\documentclass[prb,twocolumn,superscriptaddress,amsmath]{revtex4-1}
\usepackage{graphicx}
\usepackage{float}
\usepackage{color}

\usepackage{amssymb}
\usepackage{bm}

\bibpunct{[}{]}{,}{n}{}{}

\begin{document}


\title{Time-dependent reflection at the localization transition}


\author{Sergey E. Skipetrov}
\email{Sergey.Skipetrov@lpmmc.cnrs.fr}
\affiliation{Univ. Grenoble Alpes, CNRS, LPMMC, 38000 Grenoble, France}
\author{Aritra Sinha}
\affiliation{Univ. Grenoble Alpes, CNRS, LPMMC, 38000 Grenoble, France}
\affiliation{National Institute of Technology, Rourkela, Odisha, 769008, India}

\date{\today}

\begin{abstract}
A short quasi-monochromatic wave packet incident on a semi-infinite disordered medium gives rise to a reflected wave. The intensity of the latter decays as a power law $1/t^{\alpha}$ in the long-time limit. Using the one-dimensional Aubry-Andr\'{e} model, we show that in the vicinity of the critical point of Anderson localization transition, the decay slows down and the power-law exponent $\alpha$ becomes smaller than both $\alpha = 2$ found in the Anderson localization regime and $\alpha = 3/2$ expected for a one-dimensional random walk of classical particles.
\end{abstract}

\maketitle

\section{Introduction}
\label{sec_intro}

Analyzing waves (light, sound, etc.) reflected by a disordered medium is an efficient and practical way of acquiring information about the medium. Imaging and nondestructive testing in reflection mode are widely used in medical \cite{durduran10,boas10} and industrial \cite{shah14,hohler14} applications. In comparison with transmission geometry, measuring in reflection ensures a comfortable signal power (because most of the incident power is reflected for thick media) and does not require access to two opposite sides of the sample under study, which is a non-negligible practical advantage. Recent studies of fundamental wave phenomena taking place in strongly disordered media, such as, e.g., Anderson localization \cite{anderson58,lagendijk09}, also exploit the reflection geometry more and more often \cite{schuur99,johnson03,douglass11,aubry14,cobus16}. Meanwhile, most of the well-established results in this research field have been obtained for quantities measured in transmission, which can be explained by the history of the subject: many results were first established for electron scattering in disordered solids and extended to ``classical'' waves (light, sound, etc.) only later. Transmission of electrons through a disordered sample determines the electrical conductance of the latter, which is the principal physical quantity that can be measured in an experiment. Reflection measurements, if possible at all, are difficult to realize in the realm of electronics, and hence they were given little attention. Nevertheless, the reflection coefficient of a disordered sample is related to the probability for a wave to return to its initial position, the so-called return probability, which is one of the fundamental quantities in the Anderson localization theory \cite{evers08,wolfle10,cherroret08}. Therefore, reflection measurements have a potential to yield direct information about Anderson localization.

A recent work \cite{aubry14} reported measurements of the average time-dependent reflection coefficient $\langle R(t) \rangle$ of a short pulse of ultrasound that was tightly focused on a surface of a strongly disordered three-dimensional (3D) solid sample (angle brackets $\langle \ldots \rangle$ denote ensemble averaging from here on). The intensity of the reflected wave was measured at the same point where the incident pulse was focused. Depending on the central frequency of the pulse, different regimes of propagation were identified: diffuse scattering leading to $\langle R(t) \rangle \propto 1/t^{5/2}$ at long times or Anderson localization yielding $\langle R(t) \rangle \propto 1/t^{2}$. Although both these results can be understood in the frameworks of available theories \cite{white87,titov00,johnson03,douglass11,skip04,skip06}, a behavior $\langle R(t) \rangle \propto 1/t^{\alpha}$ with $\alpha \approx 1$ discovered at a critical frequency separating the frequency ranges of diffuse and localized modes (the mobility edge) turned out to be a surprise. On the one hand, a link between the temporal decay of the return probability and multifractality of critical states was proposed in the infinite disordered media \cite{chalker88,brandes96,cuevas07,kravtsov11}, but it is not clear how to extend this result to a sample's boundary. On the other hand, the fact that the power exponent $\alpha$ does not change monotonously from the diffuse value $\alpha = 5/2$ to the localized one $\alpha = 2$ when crossing the mobility edge, may hide some interesting physics. It is also curious that the power exponent $\alpha$ slightly exceeding 1 corresponds to the slowest possible decay because the time integral of $\langle R(t) \rangle$ should converge.

\begin{figure*}
\includegraphics[width=0.8\textwidth]{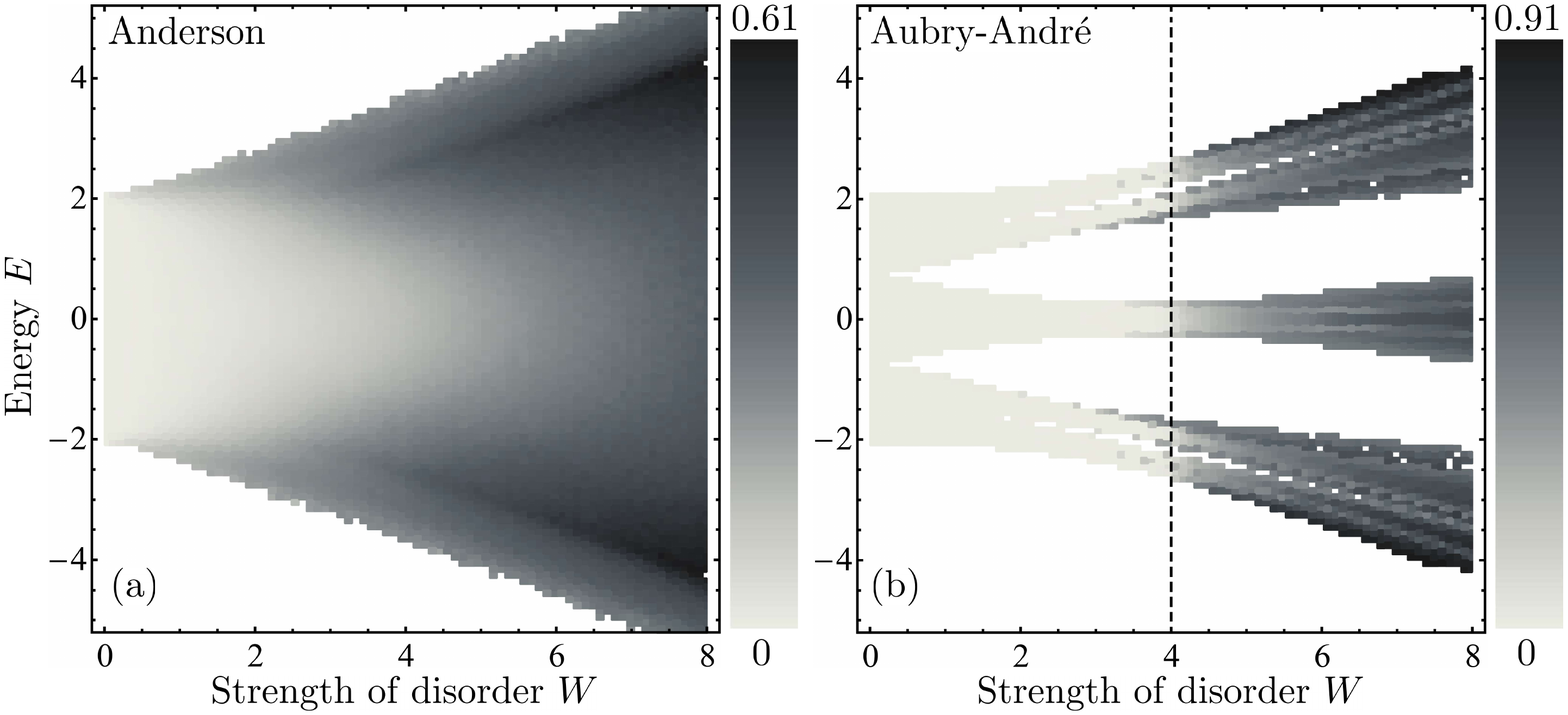}
\vspace*{-3.2cm}
\caption{\label{fig_ipr} Grayscale plot of average IPR of eigenstates of the Anderson (a) and Aubry-Andr\'{e} (b) models. Averaging was performed over 100 realizations of random potential for a system of $L = 10^3$ sites with periodic boundary conditions. The dashed vertical line in panel (b) shows the critical disorder strength $W_c = 4$ of the Aubry-Andr\'{e} model. White space corresponds to regions of parameters where no states exist. }
\end{figure*}

In the present paper, we make the first step towards understanding of the time-dependent reflection coefficient of disordered media at the critical point of localization transition by considering one of the simplest models exhibiting such a transition: the one-dimensional (1D) Aubry-Andr\'{e} model \cite{harper55,aubry80}. In this model, the random potential is quasi-periodic and is given by a deterministic formula with the randomness contained in a single parameter---the initial phase of the quasi-periodic variation. The Aubry-Andr\'{e} model and its variants have been used to study localization transitions \cite{boers07,modugno09,li17} and properties of disordered systems at the critical point \cite{tang86,siebesma87,ketz92,evan93,ketz97,rud97,evan00}, as well as to observe localization transitions in  experiments \cite{roati08,lahini09}. We will compare results obtained for this model with those for the standard Anderson model \cite{anderson58} in which the values of the potential are random and uncorrelated for different sites of a lattice. Our main result is that in the center of the energy band of the Aubry-Andr\'{e} model, the time-dependent reflection coefficient $\langle R(t) \rangle$ is roughly independent of the shape of the incident wave packet and exhibits a critical slowing down near the critical point of the Anderson localization transition. The exponent $\alpha$ of its power-law decay $\langle R(t) \rangle \propto 1/t^{\alpha}$ is below both $\alpha = 3/2$ and $\alpha = 2$ expected for classical diffusion in 1D and Anderson localization, respectively. For energies far from the band center, and especially for energies inside the energy gap of the Aubry-Andr\'{e} model, the long-time decay of $\langle R(t) \rangle$ depends on the shape of the wave packet and can be as slow as $\langle R(t) \rangle \propto 1/t^{1+\varepsilon}$, with $\varepsilon \ll 1$.

\section{Anderson and Aubry-Andr\'{e} models}
\label{model}

We want to study a wave described by the Schr\"{o}dinger equation in one dimension:
\begin{eqnarray}
i \hbar \frac{\partial}{\partial t} \psi(x,t) = \left[ -\frac{\hbar^2}{2m} \frac{\partial^2}{\partial x^2}
+ v(x) \right] \psi(x,t),
\label{schrodinger}
\end{eqnarray}
where
$m$ is the mass of the particle for which this equation provides the quantum description, and $v(x)$ is the random potential. Assuming $\psi(x,t) = \psi(x) \exp(-i \epsilon t/\hbar)$, we discretize the resulting equation for $\psi(x)$ on a lattice $x_n = n \Delta x$ applying a finite-difference approximation $(d^2/d x^2) \psi(x_n) = [\psi(x_{n+1}) - 2\psi(x_n) + \psi(x_{n-1})]/\Delta x^2$ for the second-order derivative. Setting $\hbar = 1$, $\Delta x = 1$, $2 m = 1$ (which fixes the units of energy and time) and redefining the energy $E = 2 - \epsilon$ and the potential $V_n = - v(x_n)$, we arrive at the standard tight-binding model with diagonal disorder:
\begin{eqnarray}
\psi_{n-1} + V_n \psi_n + \psi_{n+1} = E \psi_n,
\label{tight}
\end{eqnarray}
where $\psi_n = \psi(x_n)$. Assuming $V_n = 0$ and $\psi_n = A \exp(i k n)$, one obtains the free-space dispersion relation of the lattice model (\ref{tight}): $E = 2 \cos k$.

\begin{figure*}
\includegraphics[width=0.8\textwidth]{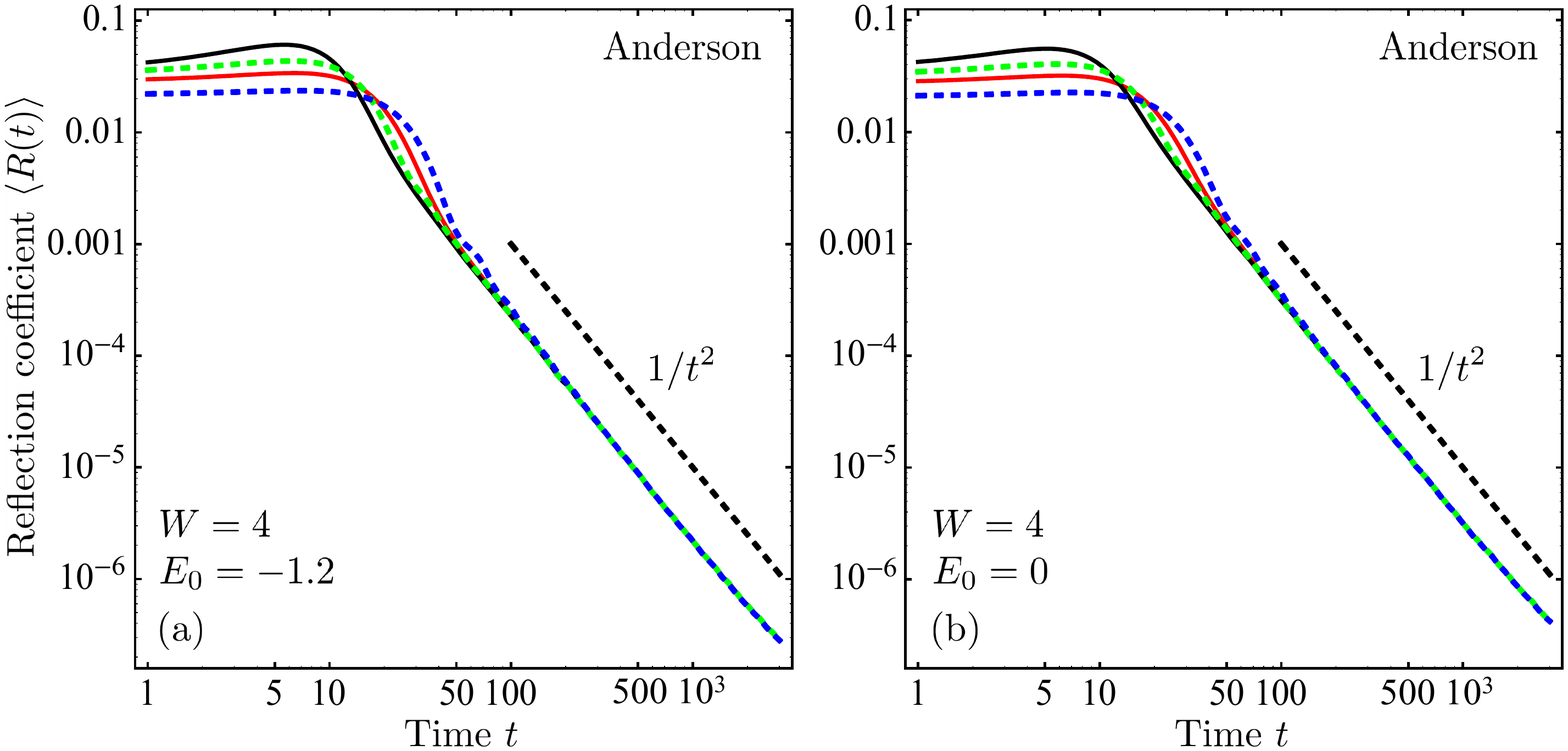}
\vspace*{-3.2cm}
\caption{\label{figanderson} Average time-dependent reflection coefficient for the Anderson model computed for a system of length $L = 10^3$, disorder strength $W=4$, and the central energy of the wave packet $E_0 = -1.2$ (a) or 0 (b). Solid and dashed lines correspond to the Gaussian [Eq.\ (\ref{fourierg})] and parabolic [Eq.\ (\ref{fourierp})] spectra, respectively, and to two different spectral widths $\sigma = 0.1$ (black and green lines) and 0.05 (red and blue lines). The results are averaged over $10^4$ independent realizations of random potential. Dashed straight lines illustrate the power-law decay at long times.}
\end{figure*}

In the following, we will compare results for two different models that are particular cases of Eq.\ (\ref{tight}). In the Anderson model with uncorrelated disorder \cite{anderson58} (for brevity referred to as simply ``Anderson model'' from here on), the on-site potentials $V_n$ are assumed to be random, uncorrelated, and uniformly distributed between $-W/2$ and $W/2$, where $W$ measures the strength of disorder. This model is extensively studied in the literature (see, e.g., Ref.\ \cite{muller11} for a pedagogical introduction, further references, and an analysis relevant to our work). In the Aubry-Andr\'{e} model \cite{harper55,aubry80}, $V_n = (W/2) \cos(2 \pi \gamma n + \phi)$, where $\gamma$ is an irrational Diophantine number and $\phi$ is a random phase uniformly distributed between 0 and $2\pi$. Without loss of generality and following some of the previous studies \cite{boers07,li17,evan00}, we set $\gamma = (\sqrt{5} - 1)/2$. Whereas the on-site potential $V_n$ varies between $-W/2$ and $W/2$ in both models, the variation is completely random for the Anderson model and quasi-periodic for the Aubry-Andr\'{e} model. This difference turns out to be of fundamental importance because the eigenstates of the Anderson model are exponentially localized in space for arbitrary, even infinitesimal disorder, whereas the eigenstates of the Aubry-Andr\'{e} model are extended for $W < W_c$ and localized for $W > W_c$. A localization transition takes place at $W = W_c = 4$ \cite{aubry80}. We illustrate this difference in Fig.\ \ref{fig_ipr} where we show the average inverse participation ratio
\begin{eqnarray}
\langle \mathrm{IPR} \rangle= \left\langle \sum_{n=1}^L |\psi_n|^4 \right\rangle
\label{ipr}
\end{eqnarray}
for a system of $L = 10^3$ sites. IPR measures the spatial localization of a state and varies from $1/L \ll 1$ for a state extended over the entire system to 1 for a state localized on a single site. The eigenenergies $E$ and eigenstates $\bm{\psi} = (\psi_1, \psi_2, \ldots, \psi_L)^T$ are obtained by numerically solving the eignevalue problem ${\hat H} \bm{\psi} = E \bm{\psi}$ for a random Hamiltonian matrix
\begin{eqnarray}
{\hat H} = \begin{bmatrix}
    V_1 &   1 &  0  & \dots &      0 &  1 \\
      1 & V_2 &  1  &     0 & \ldots &  0 \\
    \hdotsfor{6} \\
      1 &   0 & \ldots & 0 & 1 & V_L \\
\end{bmatrix}
\label{ham}
\end{eqnarray}
corresponding to Eq.\ (\ref{tight})  with periodic boundary conditions.

In agreement with previous results \cite{harper55,aubry80,boers07,modugno09,li17,
tang86,siebesma87,ketz92,evan93,ketz97,rud97,evan00}, we see that the average IPR becomes significant only for $W > W_c = 4$, clearly identifying $W_c = 4$ as a critical value of disorder for all energies. In contrast, the growth is monotonic for the average IPR computed for the Anderson model exhibiting no criticality. Another difference between the two models is that spectral gaps develop with increasing $W$ in the spectrum of the Aubry-Andr\'{e} model around $E \simeq \pm 1$, in contrast to the Anderson model for which no gaps appear.

\section{Time-dependent reflection}

We start by defining an amplitude reflection coefficient for a monochromatic wave of energy $E$. To this end, we surround a disordered region of $L-4$ sites $n = 3, 4, \ldots, L-2$ by the free space $V_1 = V_2 = V_{L-1} = V_L = 0$, so that the total number of sites is $L$. An excitation on the left from the disordered region can be represented as a sum of incident and reflected waves:
\begin{eqnarray}
\psi_n = A e^{ikn} + B e^{-ikn},\;\; n = 1, 2,
\label{left}
\end{eqnarray}
whereas only the transmitted wave exists on the right from the disordered region:
\begin{eqnarray}
\psi_n = C e^{ikn},\;\; n = L-1, L.
\label{right}
\end{eqnarray}
Here $A$, $B$, and $C$ are the amplitudes of the incident, reflected and transmitted waves, respectively, and $k = k(E)$ is the wavenumber determined according to the free-space dispersion relation. For a given energy $E$, we set $C$ equal to an arbitrary complex number (say, $C = \exp[-ik(L-1)]$), which determines $\psi_{L-1} = 1$ and $\psi_L = \exp(ik)$ via Eq.\ (\ref{right}), and then use Eq.\ (\ref{tight}) rewritten as $\psi_{n-1} = -\psi_{n+1} + (E - V_n) \psi_n$ to compute $\psi_{L-2}$, $\psi_{L-3}$, \ldots, $\psi_1$ successively by recursion on a computer \cite{note2}. Substitution of $\psi_1$ and $\psi_2$ into Eq.\ (\ref{left}) then yields a system of two linear equations for $A$ and $B$, which can be readily solved. The amplitude reflection coefficient is given by
\begin{eqnarray}
{\tilde r}(E) = \frac{B}{A} = e^{3ik} \frac{\psi_2 e^{-ik} - \psi_1}{\psi_1 e^{-ik} - \psi_2}.
\label{amplrefl}
\end{eqnarray}

\begin{figure*}
\includegraphics[width=0.8\textwidth]{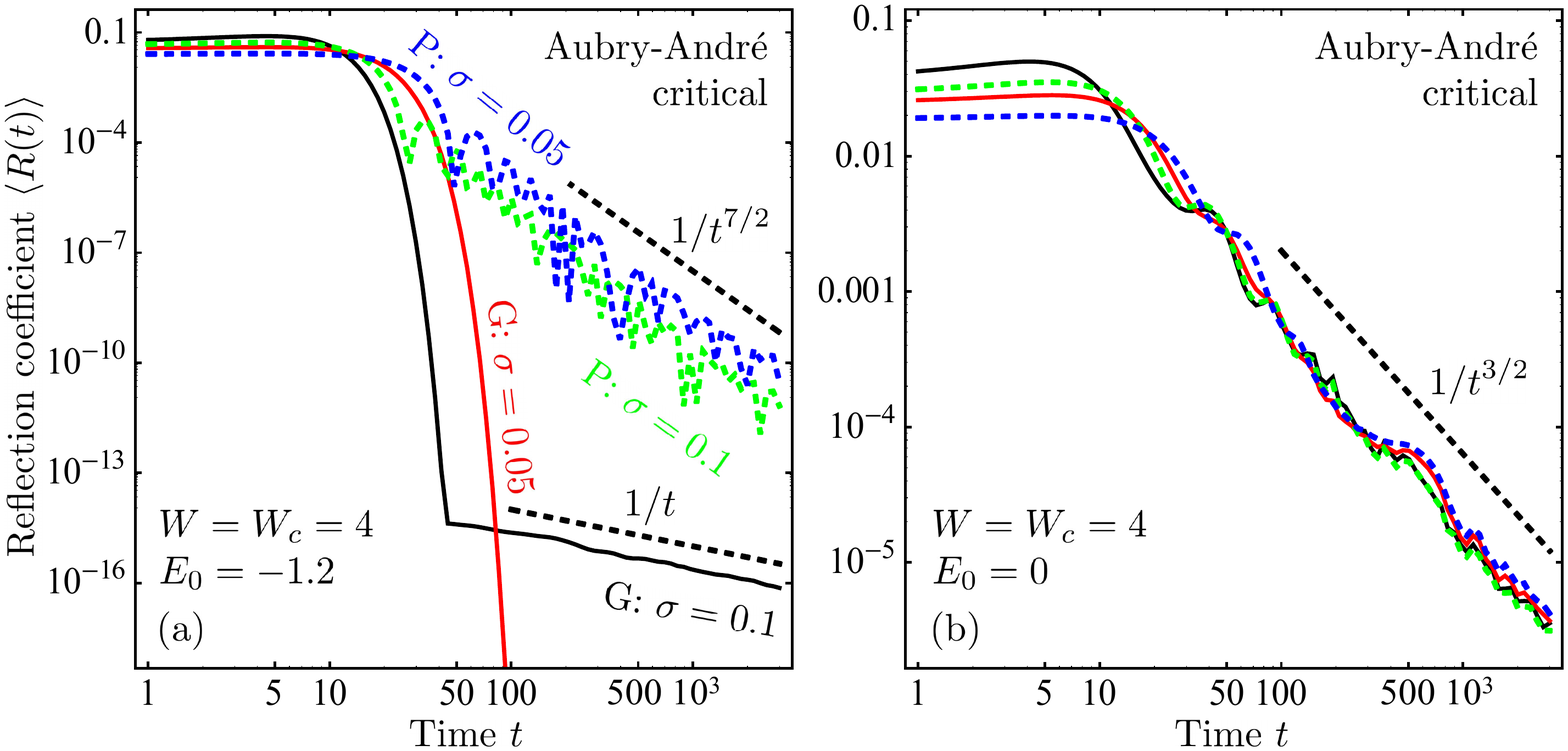}
\vspace*{-3.2cm}
\caption{\label{figaaw4} Same as Fig.\ \ref{figanderson} but for the Aubry-Andr\'{e} model. Letters `G' and `P' mark lines corresponding to the Gaussian and parabolic spectra of the incident wave packet, respectively.}
\end{figure*}

To compute the time-dependent reflection coefficient and study its sensitivity to the shape of the incident pulse, we consider wavepackets with Gaussian (G) and parabolic (P) spectra:
\begin{eqnarray}
{\tilde f_G(E)} &=& \frac{(2 \pi)^{1/4}}{\sqrt{\sigma}} \exp\left[-\frac{(E-E_0)^2}{4 \sigma^2} \right],
\label{fourierg}
\\
{\tilde f_P(E)} &=& \sqrt{\frac{15 \pi}{16 \sigma}} \left[1 - \frac{(E-E_0)^2}{4 \sigma^2} \right], \;
\left| E-E_0 \right| < 2 \sigma.\;\;\;\;\;\;\;
\label{fourierp}
\end{eqnarray}
The time profiles of these pulses are given by, respectively,
\begin{eqnarray}
f_G(t) &=& \frac{1}{2 \pi} \int\limits_{-\infty}^{\infty} dE {\tilde f_G(E)} e^{i E t}
\nonumber \\
&=& \frac{1}{(2\pi)^{1/4} \sqrt{\tau}} \exp(i E_0 t -t^2/4\tau^2),
\label{timeg}
\\
f_P(t) &=& \frac{\sqrt{30}}{2 \sqrt{\pi} \sqrt{\tau}} \left( \frac{\tau}{t} \right)^2 \left[ \frac{\sin(t/\tau)}{t/\tau} - \cos(t/\tau) \right]
e^{i E_0 t},\;\;\;\;\;\;
\label{timep}
\end{eqnarray}
where $\tau = 1/2\sigma$ is the pulse duration and
\begin{eqnarray}
\int\limits_{-\infty}^{\infty} dt |f_{P,G}(t)|^2 = \frac{1}{2\pi} \int\limits_{-\infty}^{\infty} dE |{\tilde f_{P,G}}(E)|^2 = 1.
\label{norm}
\end{eqnarray}
The reflected field is then
\begin{eqnarray}
r(t) = \frac{1}{2\pi} \int\limits_{-2}^{2} dE\; {\tilde r}(E) {\tilde f}_{P,G}(E) \exp(i E t)
\label{rt}
\end{eqnarray}
and the averaged reflected intensity
\begin{eqnarray}
\langle R(t) \rangle &=&  \langle |r(t)|^2 \rangle
\nonumber \\
&=& \frac{1}{(2\pi)^2} \int\limits_{-2}^{2} dE_1 \int\limits_{-2}^{2} dE_2
\langle {\tilde r}(E_1) {\tilde r}^*(E_2) \rangle
\nonumber \\
&\times& {\tilde f}_{P,G}(E_1) {\tilde f}_{P,G}^*(E_2)
 \exp[i (E_1 - E_2) t].
\label{reflint}
\end{eqnarray}
Note that the integrations in Eqs.\ (\ref{rt}) and (\ref{reflint}) are restricted to the energy band $-2 \leq E \leq 2$ of the homogeneous system without disorder despite the fact that states appear outside this band for $W > 0$, see Fig.\ \ref{fig_ipr}. The reflection coefficient ${\tilde r}(E)$, however, can only be defined inside the band of the homogeneous system because it characterizes the amplitude ratio of incident and reflected waves, which both exist outside the disordered region only.
The finiteness of the energy band of our model can have an important impact on the long-time behavior of the reflection coefficient $\langle R(t) \rangle$, as we will see from the following.

In our calculations we use Eq.\ (\ref{amplrefl}) to compute ${\tilde r}(E)$ for a large number ($10^4$) of random realizations of disorder $\{ V_n \}$, determine the correlation function $\langle {\tilde r}(E_1) {\tilde r}^*(E_2) \rangle$ by averaging over $\{ V_n \}$, and use Eq.\ (\ref{reflint}) to obtain the average time-dependent reflection coefficient corresponding to an incident pulse of a given central energy $E_0$, shape (parabolic or Gaussian), and bandwidth $\sigma$. In the following we set $L = 10^3$, which is long enough to consider that our disordered samples are effectively semi-infinite (see Appendix \ref{appa}), and discretize the energy $E$ with a step $\Delta E = 10^{-3}$, which is sufficient to obtain reliable results for $\langle R(t) \rangle$ up to times $t \sim 2 \pi/\Delta E \sim 5 \times 10^3$.

\section{Results}
\label{res}

To start with, we compute the time-dependent reflection coefficient for the Anderson model. Figure \ref{figanderson} shows $\langle R(t) \rangle$ for $W= 4$, two values of the central energy $E_0$ of the incident wave packet, and two values of its spectral width $\sigma$. We obtain very similar results for any $W \in [2, 8]$ and $E_0 \in [-2 + 2\sigma, 2- 2\sigma]$ as far as $\sigma \ll 1$. Anderson localization of all eigenstates at arbitrary disorder $W$ leads to a universal long-time decay $\langle R(t) \rangle \propto 1/t^2$ independent of the spectral shape of the incident wavepacket, its central energy $E_0$, and its spectral width $\sigma$. This is in agreement with expectations following from analytic theories for systems of any dimensionality \cite{white87,titov00,skip04,skip06}.

\begin{figure*}
\includegraphics[width=0.8\textwidth]{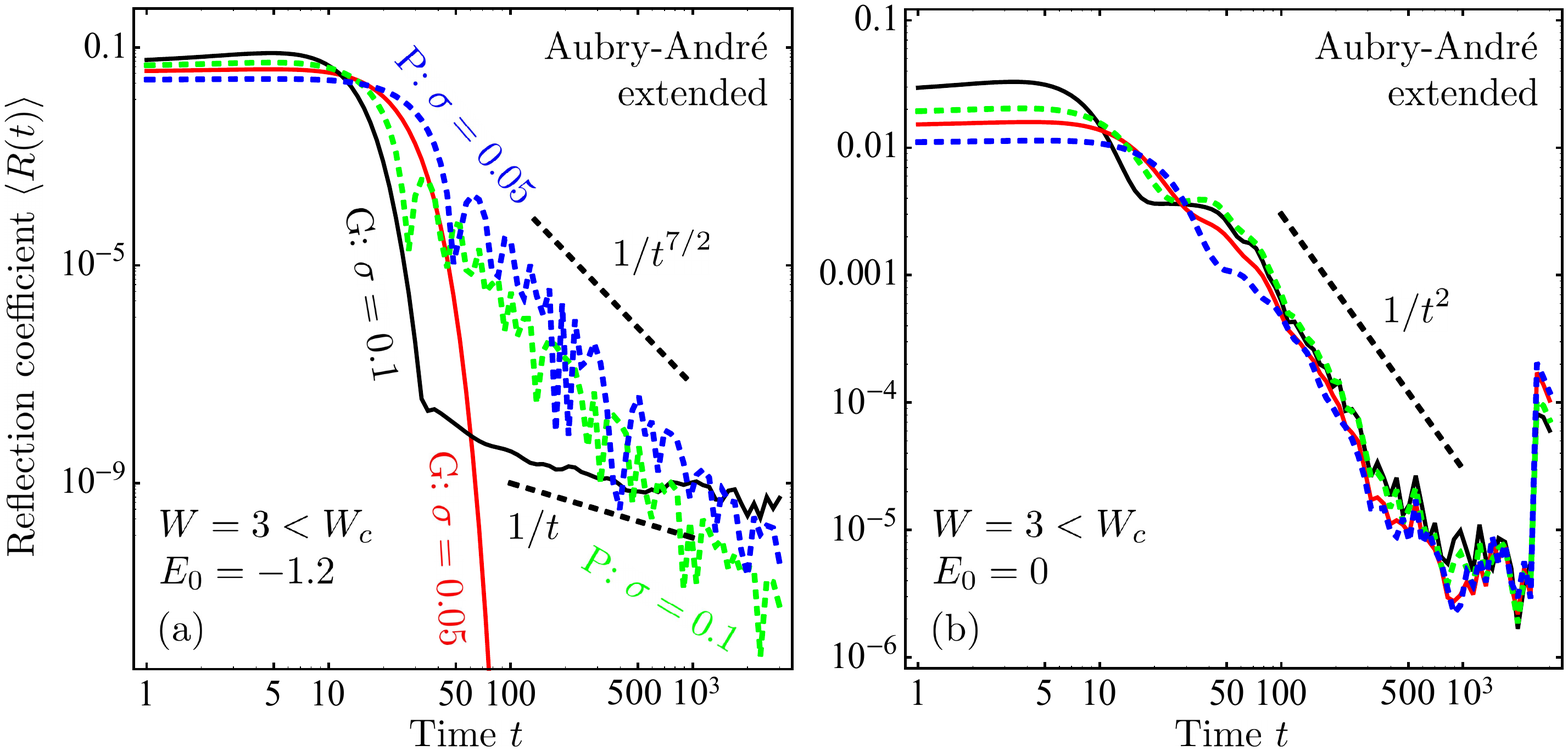}\\
\vspace*{-3.2cm}
\includegraphics[width=0.8\textwidth]{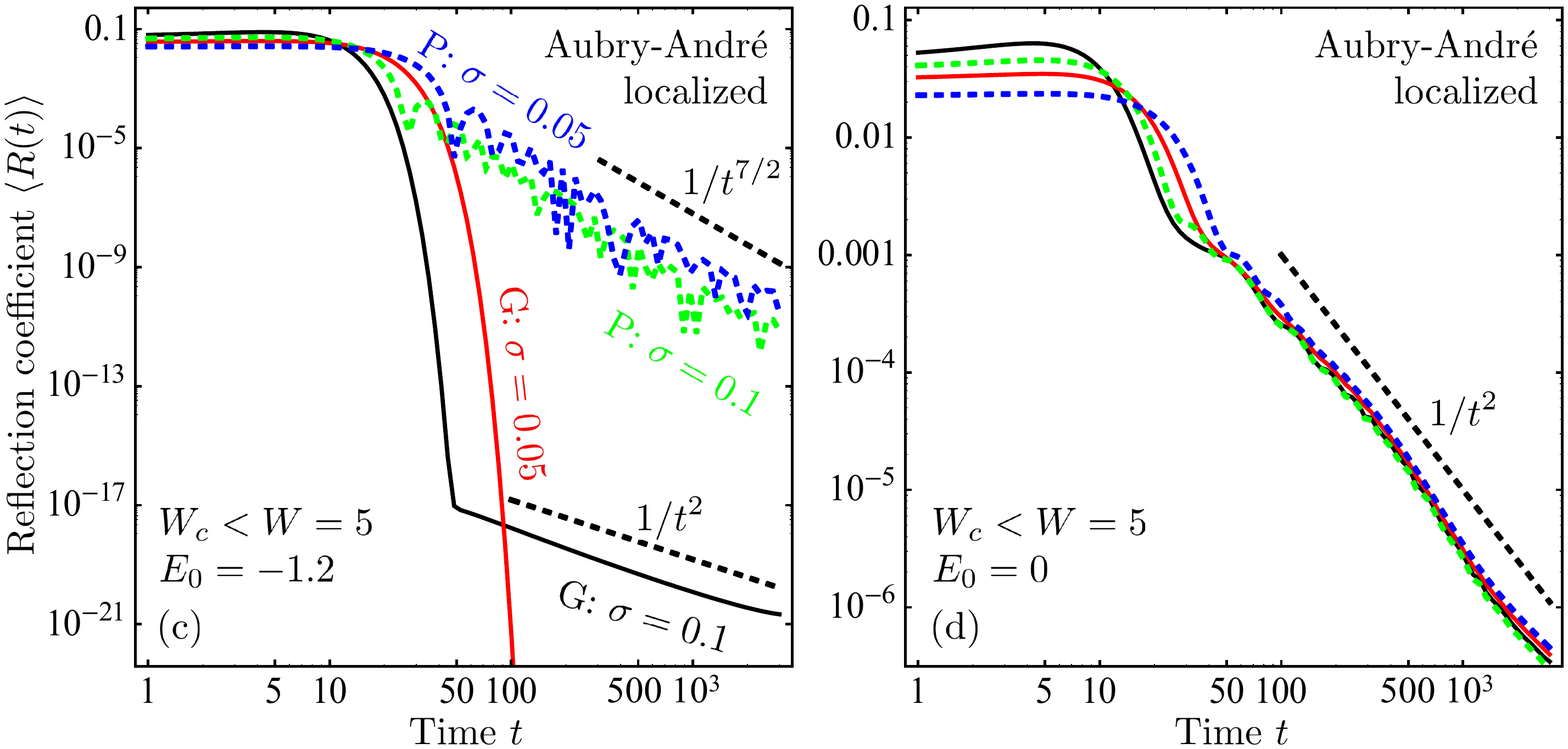}
\vspace*{-3.2cm}
\caption{\label{figaaw35} Same as Fig.\ \ref{figaaw4} but for $W = 3$ (a,b) and 5 (c,d).}
\end{figure*}

In contrast to the Anderson model, the Aubry-Andr\'{e} model yields a reflection coefficient that strongly depends on $E_0$, $\sigma$, and the shape of the incident wave packet. Figure \ref{figaaw4}(a) illustrates this for the critical value of disorder $W = W_c = 4$ and $E_0 = -1.2$ at which differences between $\langle R(t) \rangle$ corresponding to different wave packet shapes and widths $\sigma$ are maximal. The long-time decay of $\langle R(t) \rangle$ is approximately power-law in all cases, but it is clearly due to the abrupt cut of the Gaussian spectrum of the incident wave packet at the edges $E = \pm 2$ of the energy band of the model (for the Gaussian wave packet) or to the exact vanishing of the parabolic spectrum beyond $E = E_0 \pm 2 \sigma$. This is particularly obvious for the Gaussian wave packet for which dividing $\sigma = 0.1$ by 2 suppresses the effect of spectral cut by an exponentially large amount, shifting the power-law decay of $\langle R(t) \rangle$ to longer times and to much lower intensities (G: $\sigma = 0.05$ in Fig.\ \ref{figaaw4}(a); the power-law part of $\langle R(t) \rangle$ is beyond the range of the vertical axis). However, the resulting reflection coefficient cannot be interpreted simply as a specularly reflected incident wave packet, which would give $\langle R(t) \rangle \propto 1/t^2$ and $1/t^4$ for the Gaussian and parabolic spectra, respectively. Instead, the reflection coefficient shows a slower decay, which can be as slow as $\langle R(t) \rangle \propto 1/t^{1+\varepsilon}$, with $\varepsilon \ll 1$, for the Gaussian wave packet.

While for most central energies $E_0$ of the incident wave packet, the time-dependent reflection coefficient strongly depends on the shape and width of the latter, a certain degree of universality is achieved for $E_0$ around the center of the band $E = 0$, which lies far enough from the band edges $E = \pm 2$. As we show in Fig.\ \ref{figaaw4}(b), the long-time decay of $\langle R(t) \rangle$ becomes roughly independent of the incident wave packet details in this case and can be approximately described by a $1/t^{3/2}$ law. This relative universality of results obtained for $E_0 = 0$ is preserved at other values of disorder strength $W$, as we show in Fig.\ \ref{figaaw35}(b,d) for $W = 3$ and 5. In contrast, $|E_0| \sim 1$ yields nonuniversal results whatever $W$ [see Fig.\ \ref{figaaw35}(a,c)].

\begin{figure}
\hspace*{4mm}
\includegraphics[width=0.8\textwidth]{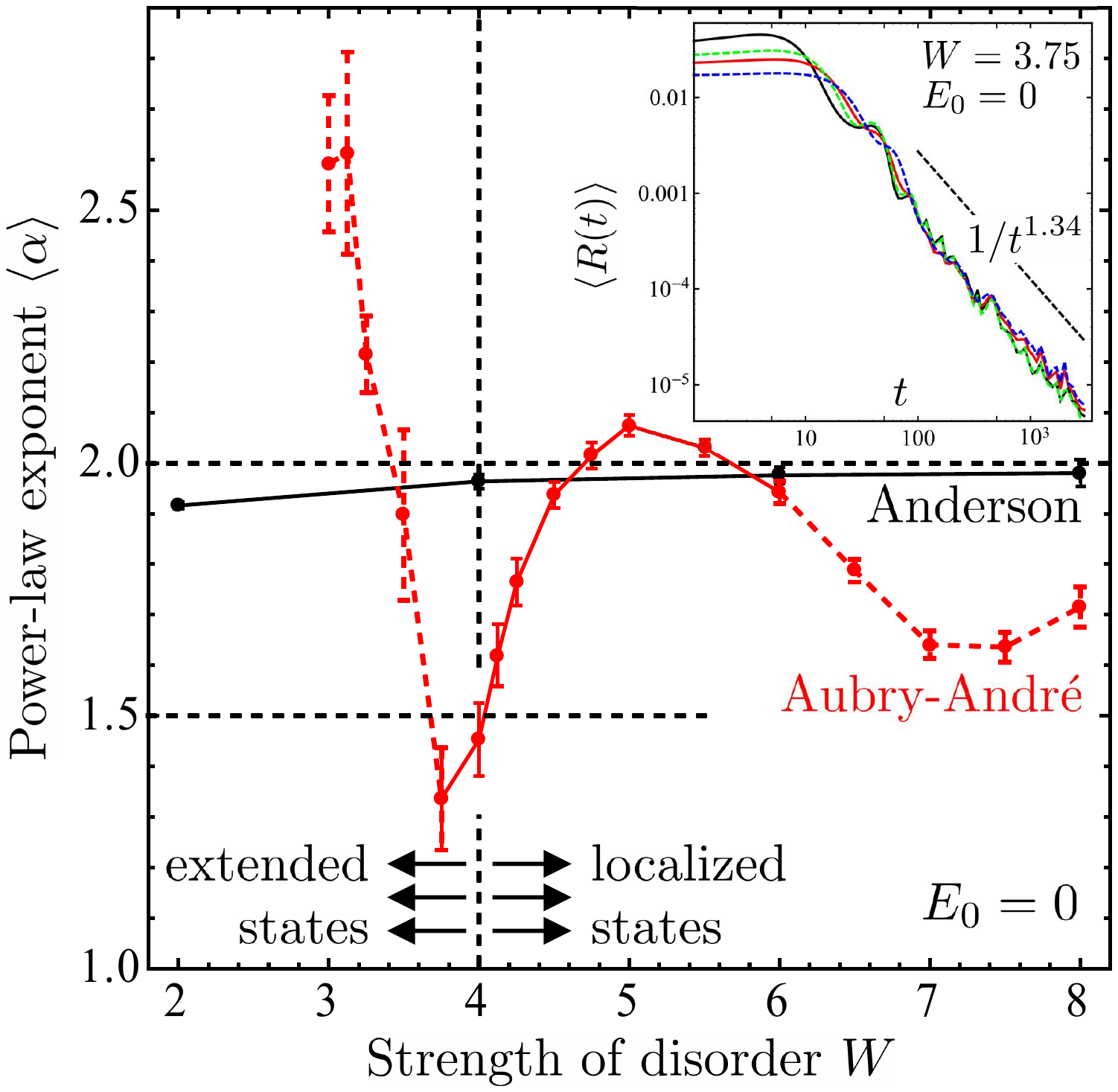}
\vspace*{-3.4cm}
\caption{\label{figexp} The power-law exponent $\langle \alpha \rangle$ of the average time-dependent reflection coefficient for the Anderson and Aubry-Andr\'{e} models obtained from the fits to the results presented in Figs.\ \ref{figanderson}(b), \ref{figaaw4}(b), \ref{figaaw35}(b,d) and similar results for other values of $W$. Long times $100 < t < 3000$ were used for the fits for all $W$ in the case of Anderson model and for $W > 3.5$ in the case of Aubry-Andr\'{e} model. For the Aubry-Andr\'{e} model with $W \leq 3.5$, only $100 < t < 500$ were used. The long-time behavior of $\langle R(t) \rangle$ for the Aubry-Andr\'{e} model starts to deviate from a pure power-law for $W < 3.5$ and $W > 6$ (shown by dashed lines). The error bars show the uncertainties due to both statistical fluctuations in the numerical data and the differences between $\alpha$ obtained for the different shapes and widths of the incident wavepacket spectrum. The vertical dashed line shows the critical point $W_c = 4$ of the Aubry-Andr\'{e} model; the two horizontal dashed lines show $\alpha = 1.5$ and 2 expected for diffuse and localized waves, respectively. Inset: The slowest decay of $\langle R(t) \rangle$ observed for $W = 3.75$.}
\end{figure}

In an attempt to systemize the results obtained in the center of the band ($E_0 = 0$), we fit the long-time behavior of the average reflection coefficient by a power law: $\ln \langle R(t) \rangle = \beta - \alpha \ln t$. The fits are performed at a fixed $W$ for the two spectral shapes of the incident pulse (\ref{fourierg}) and (\ref{fourierp}), and for the two spectral widths $\sigma = 0.05$ and 0.1 in both cases. The best-fit power exponents $\alpha$ obtained from the four fits are averaged to obtain a single exponent $\langle \alpha \rangle$. The result is shown in Fig.\ \ref{figexp} as a function of disorder strength $W$ and in comparison with a result of the same calculation performed for the Anderson model. In contrast to the Anderson model yielding $\langle \alpha \rangle \simeq 2$ with an accuracy below 5\% for any $W$, the Aubry-Andr\'{e} model gives $\langle \alpha \rangle$ that strongly depends on the strength of disorder. It reaches a minimum $\langle \alpha \rangle \simeq 1.34$ in the vicinity of the localization transition ($W \approx 3.75$ in Fig.\ \ref{figexp}; the corresponding $\langle R(t) \rangle$ are shown in the inset) although not exactly at the transition $W = W_c = 4$. This value is only slightly smaller than $\alpha = 1.5$ expected for diffusion of classical particles \cite{note3}. Whereas for $W = 4.5$--6 the behavior of the Aubry-Andr\'{e} model is similar to that of the Anderson model, the quasi-periodic nature of the potential in the Aubry-Andr\'{e} model starts to play a role at larger $W$ inducing differences with the Anderson model in which the potential is random. This is manifested in slow oscillations of $\langle R(t) \rangle$ with time, superimposed on an otherwise power-law decay, and precludes a precise determination of $\alpha$ [see Fig.\ \ref{figaaw35}(d) where such oscillations start to be visible]. In the opposite limit of weak potential ($W < 3.75$) the power-law decay of $\langle R(t) \rangle$ speeds up and eventually breaks down because the length $L = 10^3$ of the simulated system becomes insufficient to model an infinite system and the reflection of the coherent part of the wave packet from the other end of the system starts to be visible in the reflected signal, as we explain in Appendix \ref{appa} [see a sharp jump of $\langle R(t) \rangle$ at $t \simeq 2.5 \times 10^3$ in Fig.\ \ref{figaaw35}(b)].

\section{Discussion}
\label{dis}

Dynamics of a wave packet in an infinite quantum or wave system at the critical point of a localization transition can be related to the multifractal properties of critical eigenstates \cite{chalker88,brandes96,cuevas07,kravtsov11}. Such a relation has been worked out for the critical 1D Aubry-Andr\'{e} (or Harper) model long time ago \cite{siebesma87,ketz92,evan93,ketz97}. In particular, the probability for a particle to return to a given lattice site after a long time $t$ (the so-called return probability) is predicted to decay very slowly as a power law $P(t) \propto 1/t^{\alpha}$ with $\alpha \simeq 0.14$ \cite{ketz92,evan93}. The analysis presented above shows that this behavior is significantly modified at a boundary of an open disordered system. Indeed, for a sufficiently narrow wave packet, the average reflection coefficient $\langle R(t) \rangle$ is expected to exhibit the same time dependence as the return probability to the first site of the disordered region ($n = 3$ in our notation). However, for energy conservation reasons, $\langle R(t) \rangle$ cannot decay slower the $1/t^{1+\varepsilon}$, with arbitrary small but positive $\varepsilon$. This is due to the requirement of convergence for the integral of $\langle R(t) \rangle$ over time $t$. In principle, a slower decay of $\langle R(t) \rangle$ might be possible in a limited time range $t < t_{\mathrm{cutoff}}$ after which a faster (power-law or exponential) decay would take over, but we do not find any sign of such a behavior in our calculations.

It may be tempting to attribute the difference in the time decay of $P(t)$ expected in an infinite medium and the calculated decay of $\langle R(t) \rangle$ to the fact that, by construction, the latter is determined at a boundary of a disordered system. Taking into account the impact of the boundary on the multifractality might then cure the problem. Such an approach may indeed be justified for a weakly open system of finite size in which the eigenstates of a closed system acquire decay rates $\Gamma$ which are much smaller than the typical spacing $\Delta$ between adjacent energy levels of a closed system. However, in our calculations $\Delta = 4/L = 4 \times 10^{-3}$ and the reflected signal decays by several orders of magnitude already for times $t < 2\pi/\Delta \simeq 1500$. This witnesses that $\Gamma \gg \Delta$ for the majority of quasi-states contributing to the reflected signal (the prefix ``quasi-'' reflects the fact that the states have acquired finite lifetimes $1/\Gamma$). In such a situation, the physical processes dominating the decay of $\langle R(t) \rangle$ are fundamentally different from those responsible for the slow decay of the return probability $P(t)$ in an infinite (or finite but closed) system. Indeed, the wave dynamics in the infinite system is determined by free oscillations of its eigenstates. The eigestates are excited at $t=0$ and then does not decay in time. The average result of their superposition is governed by the correlation of intensities of different eigenstates \cite{chalker88,brandes96,cuevas07,kravtsov11}. This correlation, in its turn, is sensitive to the multifractal structure of the states, which explains the physical mechanism behind the link between the decay of $P(t)$ and the multifractality.

The situation is drastically different in an open disordered system, where quasi-states have finite lifetimes $1/\Gamma \ll 1/\Delta$ because of strong energy leakage to the outside world. The reflection coefficient measures precisely this leakage. The intensities of quasi-states decay exponentially in time as $\exp(-\Gamma t)$, with random decay rates $\Gamma$ \cite{kottos05}. Weights of different quasi-states contributing to $\langle R(t) \rangle$ are correlated with their decay rates because the states that leak the most are also the most efficiently excited by the incident wave packet. $\langle R(t) \rangle$ is obtained as an integral of $\exp(-\Gamma t)$ multiplied by a weight $P(\Gamma)$ of states with a decay rate $\Gamma$ \cite{skip04}. In the localized regime, for example, $P(\Gamma) \propto \Gamma$ and $\langle R(t) \rangle \propto 1/t^2$ as we see from Fig.\ \ref{figanderson}. $P(\Gamma) \propto \Gamma^{\alpha-1}$ would yield $\langle R(t) \rangle \propto 1/t^{\alpha}$. Even though the precise link between the multifractality and $P(\Gamma)$ at the critical point of the localization transition is not clear at the moment, it can be studied with help of numerical approaches similar to those used in this paper. Such a study is, however, outside the scope of the present work.

Another feature of the Aubry-Andr\'{e} model that might, in principle, affect the time-dependent reflection coefficient, is the presence of localized boundary states originating from the nontrivial topological properties of the model. In a system with closed boundaries, these states appear in the spectral gaps opening around $E \sim \pm 1$ \cite{kraus12} for both $W < W_c$ and $W \geq W_c$. A weak opening of the boundaries should confer to these states a finite lifetime that we expect to be shorter than the lifetime of modes exploring the bulk of the system. Indeed, because of their localization at the boundary, the boundary states are likely to couple to the outside world more efficiently than states that have only a part of their weight near a boundary. As a result the boundary states may affect the short-time behavior of $\langle R(t) \rangle$ but are unlikely to play any role at long times. In addition, their impact should be visible at any $W$ and not only at $W = W_c$ in which we are mainly interested here. For a system that is fully open, the above arguments become even stronger and it is even unclear if any signature of localized boundary states may remain in $\langle R(t) \rangle$. We thus conclude that the localized boundary states inherent for the Aubry-Andr\'{e} model with closed boundaries, should not affect the long-time behavior of the reflection coefficient $\langle R(t) \rangle$ in a model with open boundaries that we consider in this work.

\section{Conclusion}

The 1D Aubry-Andr\'{e} model yields the average time-dependent reflection coefficient $\langle R(t) \rangle$ that strongly depends on the shape, the central energy $E_0$, and the spectral width $\sigma$ of the incident wave packet, except in the center of the band $E_0 = 0$, where roughly universal results can be obtained. In the center of the band, the long-time decay of $\langle R(t) \rangle$ is power-law: $\langle R(t) \rangle \propto 1/t^{\alpha}$. The exponent $\alpha \simeq 2$ in the localized regime $4.5 < W < 6$. Weak but visible oscillations are superimposed on the power-law decay of $\langle R(t) \rangle$ at stronger disorder $W \gtrsim 6$, making it difficult to determine $\alpha$ precisely. $\langle R(t) \rangle$ exhibits a critical slowing down in the vicinity of the critical point $W = W_c=4$, where $\alpha \simeq 1.5$. The minimum value $\alpha \simeq 1.34$ is reached slightly below the critical point at $W \simeq 3.75$.

When the central energy of the incident wave packet $E_0$ is far from the center of the band and, in particular, when $E_0$ is inside one of the spectral gaps that open around $E \simeq \pm 1$, the long-time decay of  $\langle R(t) \rangle$ depends on the shape and the spectral width of the incident pulse and, for a Gaussian wave packet at the critical point $W = W_c$, can be as slow as $1/t^{\alpha}$ with $\alpha = 1 + \varepsilon$ and $\varepsilon \ll 1$. This suggests that nontrivial behaviors may result from an interplay of criticality with the band structure of a disordered system. Indeed, the interplay between Anderson localization and band gap formation has been recently shown to give rise to interesting physics in 2D disordered structures as well \cite{froufe17}. Such an interplay may also be at the origin of the slow decay of time-dependent reflection $\langle R(t) \rangle \propto 1/t^{\alpha}$ with $\alpha \simeq 1$ observed in the experiments at a critical point of localization transition that happened to fall near an edge of a spectral gap of a 3D disordered sample \cite{aubry14}. Further studies are needed to explore this conjecture in more detail as well as to understand the role of dimensionality (1D versus 2D and 3D) in this context.

\begin{acknowledgments}
A.S. thanks LPMMC for hosting him during the summer 2017. This work was supported by the Agence Nationale de la Recherche under Grant No. ANR-14-CE26-0032 LOVE. All the computations presented in this paper were performed using the Froggy platform of the CIMENT infrastructure ({\tt https://ciment.ujf-grenoble.fr}), which is supported by the Rhone-Alpes region (grant CPER07\verb!_!13 CIRA) and the Equip@Meso project (reference ANR-10-EQPX-29-01) of the programme Investissements d'Avenir supervised by the Agence Nationale de la Recherche.
\end{acknowledgments}

\appendix

\section{Finite-size effects}
\label{appa}

\begin{figure*}
\includegraphics[width=0.8\textwidth]{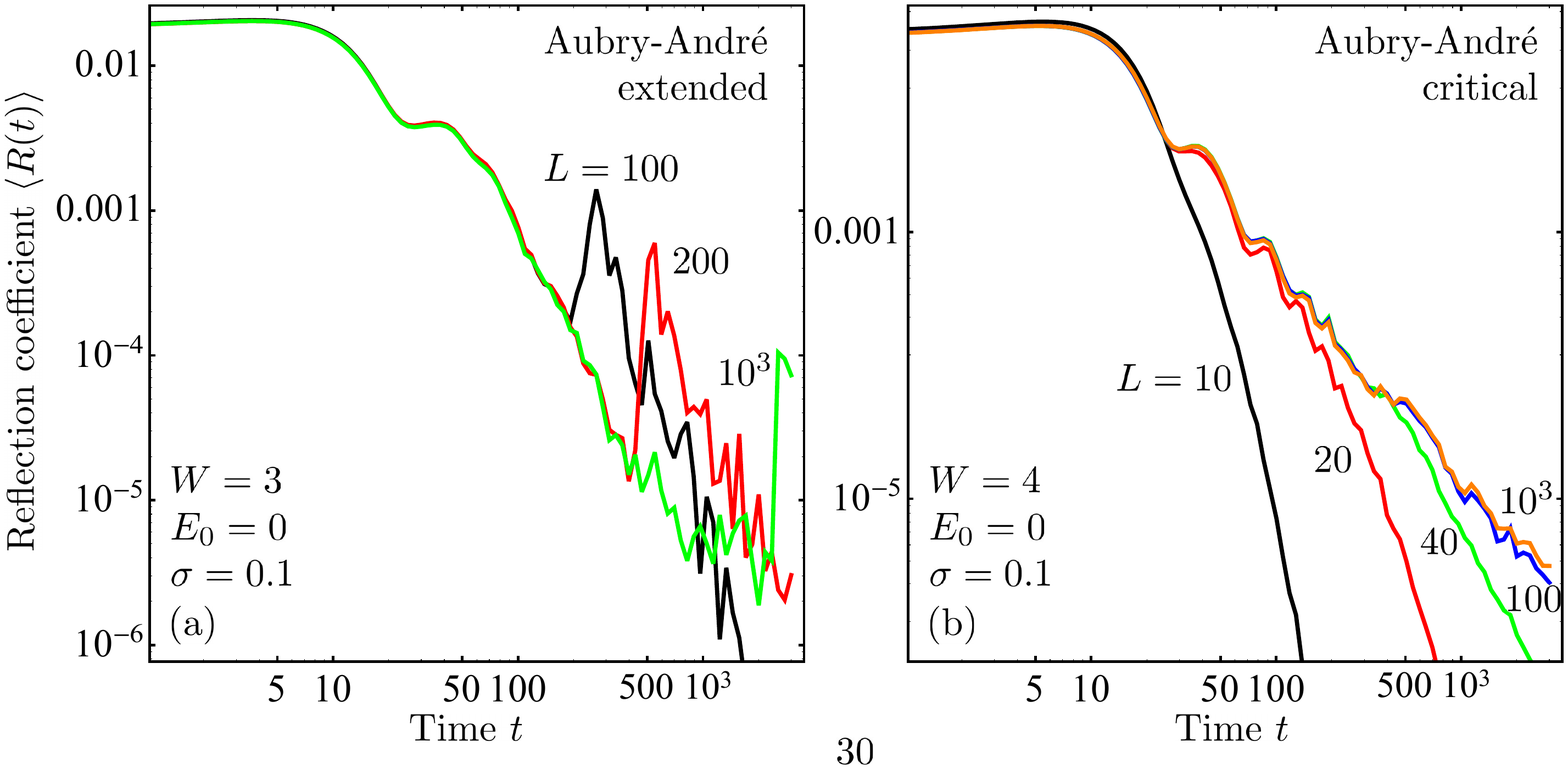}
\vspace*{-3.2cm}
\caption{\label{figfs} Illustration of finite size effects in the average reflection coefficient $\langle R(t) \rangle$ of the Aubry-Andr\'{e} model in the extended (a) and critical (b) regimes. Different lines correspond to different system lengths $L = 100$, 200, $10^3$ in the panel (a) and $L = 10$, 20, 40, 100, $10^3$ in the panel (b). The incident wave packet is assumed to have the parabolic shape (\ref{timep}).}
\end{figure*}

The calculations presented in the main text have been performed for a 1D system of $L = 10^3$ sites that was assumed to model a semi-infinite system. In this Appendix we justify this assumption and show how finite-size effects manifest themselves in systems of shorter lengths.

When the strength $W$ of the quasi-periodic potential is weak ($W \lesssim 3$), the finite size of the system produces a peak in $\langle R(t) \rangle$ shortly after $t = 2L$, as we show in Fig.\ \ref{figfs}(a). This is due to the reflection of the coherent attenuated wave from the opposite end of the system and its propagation back to the beginning of the system. Remnants of this phenomenon survive even for $L = 10^3$ and are seen in Fig.\ \ref{figaaw35}(b). To ensure that it does not influence our results, we use only the numerical data corresponding to $100 \leq t \leq 500$ for our fits when $W < 3.75$, whereas times $100 \leq t \leq 3 \times 10^3$ are used when $W \geq 3.75$.

At $W \gtrsim 3.25$, the peak in $\langle R(t) \rangle$ due to the coherent signal reflected from the opposite end of the system is not visible any more (at least, for times $t \leq 3 \times 10^3$) and the finite size of the system manifests itself by a faster decay of $\langle R(t) \rangle$ after a certain time depending on $L$ but significantly exceeding $2L$, see Fig.\ \ref{figfs}(b). We attribute this decay to the leakage of wave energy out of the system at the transmission side. As follows from Fig.\ \ref{figfs}(b), for $t \leq 3 \times 10^3$ this phenomenon can be safely ignored if $L > 100$ because $\langle R(t) \rangle$ obtained for $L = 100$ and $10^3$ virtually coincide. Hence, $L = 10^3$ used in the main text is sufficient to model the behavior of a semi-infinite system for $t \leq 3 \times 10^3$.


\begin{thebibliography}{99}
\bibitem{note2}
Our computer code is written in {\tt FORTRAN} and uses quadruple precision arithmetics.
\bibitem{note3}
The reflection coefficient for diffusing classical particles can be readily established by using the 1D diffusion kernel $P(x, x', t) = \exp[-(x-x')/4Dt]/\sqrt{4 \pi Dt}$ with a source of particles at $x' \gtrsim 1$ (a point where the incident flux of ballistic particles is converted into diffusing random walkers) and applying the standard boundary condition of zero density of diffusing particles at an extrapolated boundary $x = -x_0$ with $x_0 \sim 1$. Using the image method to obey the boundary condition, we obtain $\langle R(t) \rangle \propto P(0, x', t) - P(0, -x' -2 x_0, t) \propto 1/t^{3/2}$ for $t \to \infty$.
\end{thebibliography}
\end{document}